\documentclass[11pt,a4paper]{article}
\pdfoutput=1

\usepackage{jheppub}

\usepackage{slashed}
\usepackage{ifpdf,colortbl}

\ifx\pdfoutput\undefined
\usepackage[dvips,bookmarks=false]{hyperref}	
\else
\usepackage{hyperref}	
\fi
\hypersetup{colorlinks,bookmarksopen,bookmarksnumbered,citecolor=verdes,
linkcolor=blus,pdfstartview=FitH,urlcolor=rossos}
\def\hhref#1{\href{http://arxiv.org/abs/#1}{#1}} 

\usepackage{amsfonts}

\newcommand{\abs}[1]{\left| #1 \right|}

\newcommand{\beq}{\begin{equation}}
\newcommand{\eeq}{\end{equation}}
\newcommand{\fig}[1]{~\ref{fig:#1}}

\newcommand{\One}{\hbox{1\kern-.24em I}}

\newcommand{\GeV}{\,{\rm GeV}}
\newcommand{\TeV}{\,{\rm TeV}}

\newcommand{\eq}[1]{~{\rm (\ref{eq:#1})}}

\newcommand{\Ord}{{\cal O}}

\def\Black{}
\def\Blue{}

\newcommand{\lascia}[1]{}
\makeatletter
\def\art{\@ifnextchar[{\eart}{\oart}}
\def\eart[#1]#2#3#4#5#6{{\rm #2}, {#3 #4} {\rm (#6) #5} [arXiv:{\hhref{#1}}]}
\def\hepart[#1]#2{{\rm #2, arXiv:\hhref{#1}}}
\newcommand{\oart}[5]{{\rm #1}, {#2 #3} {\rm (#5) #4}}

\newcounter{alphaequation}[equation]
\def\thealphaequation{\theequation\hbox to
0.6em{\hfil\alph{alphaequation}\hfil}}
\def\eqnsystem#1{
\def\@eqnnum{{\rm (\thealphaequation)}}
\def\@@eqncr{\let\@tempa\relax \ifcase\@eqcnt \def\@tempa{& & &} \or
  \def\@tempa{& &}\or \def\@tempa{&}\fi\@tempa
  \if@eqnsw\@eqnnum\refstepcounter{alphaequation}\fi
\global\@eqnswtrue\global\@eqcnt=0\cr}
\refstepcounter{equation} \let\@currentlabel\theequation \def\@tempb{#1}
\ifx\@tempb\empty\else\label{#1}\fi
\refstepcounter{alphaequation}
\let\@currentlabel\thealphaequation
\global\@eqnswtrue\global\@eqcnt=0 \tabskip\@centering\let\\=\@eqncr
$$\halign to \displaywidth\bgroup \@eqnsel\hskip\@centering
$\displaystyle\tabskip\z@{##}$&\global\@eqcnt\@ne
\hskip2\arraycolsep\hfil${##}$\hfil& \global\@eqcnt\tw@\hskip2\arraycolsep
$\displaystyle\tabskip\z@{##}$\hfil
\tabskip\@centering&\llap{##}\tabskip\z@\cr}

\def\endeqnsystem{\@@eqncr\egroup$$\global\@ignoretrue} \makeatother

\def\diag{\mathop{\rm diag}}

\def\Ord{{\cal O}}
\def\Lag{{\cal L}}
\def\SU{{\rm SU}}

\def\circa#1{\,\raise.3ex\hbox{$#1$\kern-.75em\lower1ex\hbox{$\sim$}}\,}

\usepackage{multicol}
\usepackage{color}
\definecolor{rosso}{cmyk}{0,1,1,0.4}
\definecolor{rossos}{cmyk}{0,1,1,0.55}
\definecolor{rossoc}{cmyk}{0,1,1,0.2}
\definecolor{blu}{cmyk}{1,1,0,0.3}
\definecolor{blus}{cmyk}{1,1,0,0.6}
\definecolor{bluc}{cmyk}{1,1,0,0.1}
\definecolor{verde}{cmyk}{0.92,0,0.59,0.25}
\definecolor{verdec}{cmyk}{0.92,0,0.59,0.15}
\definecolor{verdes}{cmyk}{0.92,0,0.59,0.4}
\definecolor{grigio}{cmyk}{0,0,0,0.07}
\definecolor{rosa}{cmyk}{0,0.1,0.1,0.02}
\definecolor{rosino}{cmyk}{0,0.05,0.05,0.02}
\definecolor{rosas}{cmyk}{0,0.3,0.25,0.05}
\definecolor{celeste}{cmyk}{0.1,0,0,0.02}
\definecolor{giallino}{cmyk}{0,0,0.4,0.02}
\definecolor{rosso}{cmyk}{0,1,1,0.4}
\definecolor{rossos}{cmyk}{0,1,1,0.55}
\definecolor{rossoc}{cmyk}{0,1,1,0.2}
\definecolor{blu}{cmyk}{1,1,0,0.3}
\definecolor{bluc}{cmyk}{1,1,0,0.1}
\definecolor{blucc}{cmyk}{0.7,0.5,0,0}
\definecolor{viola}{cmyk}{0,1,0,0.6}
\definecolor{viola2}{cmyk}{0,1,0.2,0.6}
\definecolor{verde}{cmyk}{0.92,0,0.59,0.25}
\definecolor{verdec}{cmyk}{0.92,0,0.59,0.15}
\definecolor{verdes}{cmyk}{0.92,0,0.59,0.4}
\definecolor{verdino}{cmyk}{0.12,0,0.09,0.05}
\definecolor{giallo}{cmyk}{0,0,1,0}
\definecolor{gialloverde}{cmyk}{0.44,0,0.74,0}

\font\tenrsfs=rsfs10 at 12pt

\font\sevenrsfs=rsfs7
\font\fiversfs=rsfs5
\newfam\rsfsfam
\textfont\rsfsfam=\tenrsfs
\scriptfont\rsfsfam=\sevenrsfs
\scriptscriptfont\rsfsfam=\fiversfs
\def\mathscr#1{{\fam\rsfsfam\relax#1}}

\def\Lag{\mathscr{L}}

\title{Anthropic solution to the magnetic muon anomaly: the charged see-saw}

\author[a,b]{Kristjan Kannike,}
\author[b]{Martti Raidal,}
\author[a]{David M. Straub}
\author[b,c]{and Alessandro Strumia}

\affiliation[a]{Scuola Normale Superiore and INFN, \\
Piazza dei Cavalieri 7, 56126 Pisa, Italia}
\affiliation[b]{National Institute of Chemical Physics and Biophysics, \\
Ravala 10, Tallinn, Estonia}
\affiliation[c]{Dipartimento di Fisica dell'Universit{\`a} di Pisa and INFN, \\ Italia}

\emailAdd{kristjan.kannike@sns.it}
\emailAdd{martti.raidal@cern.ch}
\emailAdd{david.straub@sns.it}
\emailAdd{astrumia@mail.df.unipi.it}

\abstract{We present models of new physics that can explain the muon $g-2$ anomaly in accord with the assumption that the only scalar existing at the weak scale is the Higgs, as suggested by anthropic selection. Such models are dubbed ``charged see-saw'' because the muon mass term is mediated by heavy leptons. The electroweak contribution to the $g-2$ gets modified by order one factors, giving an anomaly of the same order as the observed hint, which is strongly correlated with a modification of the Higgs coupling to the muon.}

\arxivnumber{1111.2551}

\begin{document}

\maketitle


\section{Introduction}
The Higgs mass naturalness problem and the observed deviation 
of the muon anomalous magnetic moment $a_\mu$~\cite{Bennett:2006fi} from its SM prediction~\cite{Davier:2010nc} 
has attracted a lot of attention. (See \cite{Jegerlehner:2009ry,Czarnecki:2001pv,Strumia:2008qa} for thorough reviews of the theory situation and possible new physics solutions.)
Supersymmetric models could nicely explain both issues.
But  the recent negative results of the LHC searches cast  doubts on the naturalness of models like the CMSSM~\cite{Strumia:2011dv,Ellwanger:2011mu,Altarelli:2011vt,Papucci:2011wy}.
and disfavor their sparticle spectra preferred by the $a_\mu$ anomaly within global fits~\cite{Farina:2011bh,Bertone:2011nj,Buchmueller:2011sw}.
This experimental situation, together with the discovery of an unnaturally small cosmological constant~(see \cite{Larson:2010gs} and refs. therein), 
suggests that the hierarchies between the cosmological, weak and gravitational scales 
could be due to anthropic selection~\cite{Weinberg:1987dv,Agrawal:1997gf}.
While there has been considerable interest  in this possibility~\cite{ArkaniHamed:2004fb,Hall:2007ja},
it is unclear if it can have testable implications.


\newpage

In this paper we assume that:
\begin{itemize}
\item[a)] The $a_\mu$ anomaly~\cite{Bennett:2006fi,Davier:2010nc}
\beq \Delta a_\mu^{\rm exp} = a_\mu^{\rm exp} - a_\mu^{\rm SM} \approx (2.8 \pm 0.8)~10^{-9}\eeq
is a real signal of new physics.
This means that new light particles must exist not far away from the weak scale.

\item[b)] The Higgs mass is small due to anthropic selection.
This presumably means that the only elementary scalar at the weak scale is the Higgs doublet:
no new elementary scalars are light, because anthropic selection would not demand the
fine-tuning necessary for their lightness.
For example, the discovery of a massive $Z'$ boson at LHC would speak against 
the anthropic scenario.
\end{itemize}
We address the following issue: are a) and b) compatible?


We will show that new fermions at the weak scale mixed with the muon are technically natural and
can explain the  $a_\mu$ anomaly provided that they realize ``charged see-saw models'', namely models
that mediate the muon mass term, because it has the same
chiral structure as the muon magnetic moment.
The new physics contribution 
is naturally of the same order as the SM electroweak contribution~\cite{Brodsky:1966mv,Jackiw:1972jz,Bars:1972pe}
\beq \Delta a_\mu^\text{SM-EW} =\frac{m_\mu^2}{(4\pi v)^2}\left( 1-\tfrac{4}{3}s_{\rm W}^2 + \tfrac{8}{3} s_{\rm W}^4 \right)
\approx 2 \times 10^{-9}\eeq
and consequently of the same order as the observed anomaly.


In section~\ref{see-saw} we classify the ``charged see-saw models'', where new heavy leptons contribute to the muon mass.
In section~\ref{g-2} we derive generic formul\ae{} for one-loop corrections to the muon magnetic moment, taking into account that in
such modes the chirality flip can be enhanced by heavy fermion masses.
In section~\ref{results} we present our results, and in section~\ref{concl} our conclusions.

\begin{figure}[t]
\centering
\includegraphics[width=0.85\textwidth]{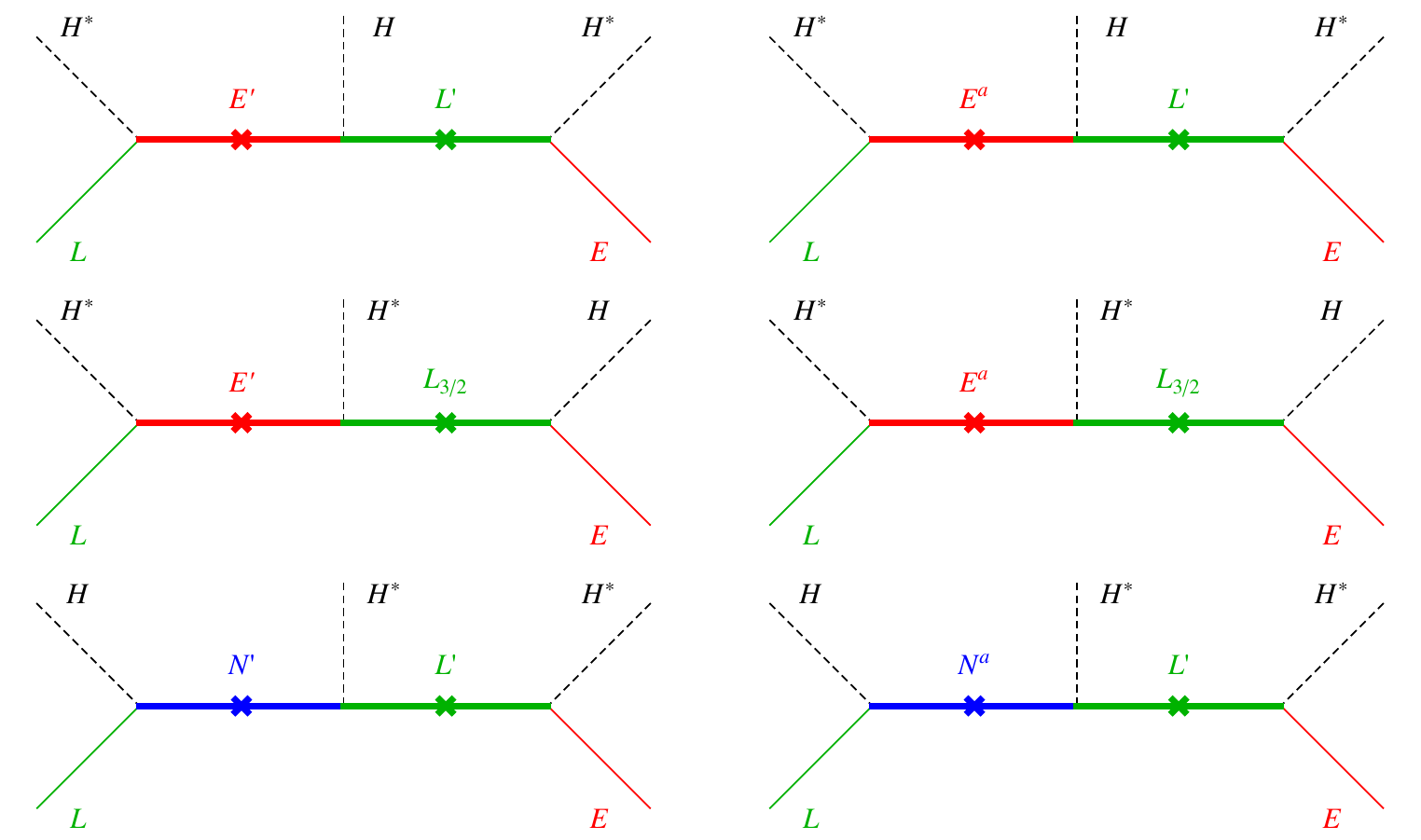}
\caption{ See-saw models that can mediate at tree level the charged lepton mass operator ${\cal O}_{HHH}$.
New heavy leptons are plotted as thick lines and listed in table~\ref{tab:listf}. The bottom-left diagram is zero (see text).} 
\label{fig:FeynTree}
\end{figure}

\begin{table}[t]
$$\begin{array}{|ccccccc|}\hline
\rowcolor[gray]{0.8}
\hbox{Name}&{\rm U}(1)_Y&\SU(2)_L&\SU(3)_{\rm c} & \hbox{$Q=T_3+Y$} & \hbox{Lepton number}& \hbox{couplings} \cr \hline\hline
\rowcolor{verdino} L' & -{1 \over 2} & 2 &1 & 0, -1  &+1 & EL'H^*  \cr
\rowcolor{verdino}L_{3/2}   &-{3\over 2} & \bar 2 &1 &-1,-2&+1&E(L_{3/2}\epsilon H)\cr 
\rowcolor{rosa} E'  &\phantom{-}1 & 1 &1 & 1&-1& E'LH^*  \cr  
\rowcolor{rosa} E^a   &\phantom{-}1 & 3 &1 &0,1,2& -1&E^a(H^* \tau^a L)\cr 
 \rowcolor{celeste}N^a  &\phantom{-}0 & 3 &1 &-1,0,+1& -1& N^a (H\epsilon \tau^a L)\cr 
 \rowcolor{celeste} N'  &\phantom{-}0 & 1 &1 & 0 &-1& N' LH \cr 
\hline\end{array}$$
\caption{\label{tab:listf} List of new leptons that can couple to the SM lepton doublet
$L = (\nu_\mu,\mu_L)$ or singlet $E=\mu_R$ 
(with the same gauge quantum numbers as $L'$ and $E'$) and to the Higgs doublet $H=(0,v+h/\sqrt{2}) $ (an SU(2) doublet with $Y=1/2$).
For each new complex field we also add the corresponding conjugate representation:
e.g.\ $L'$ is accompanied by $\bar L'$ in the $\bar 2$ representation with hypercharge $+1/2$.
}
\end{table}

\section{Charged see-saw models}\label{see-saw}
The muon  magnetic moment operator has the same  chiral structure as the muon mass operator, which in the SM is given by
$m_\mu =m_\mu^H \equiv \lambda_\mu v$ where $\lambda_\mu$ is the Yukawa coupling in the Lagrangian term
\beq-\lambda_\mu LEH^*,\eeq 
 $E $ is the right-handed muon; $L$ is the left-handed muon doublet and $H$ is the Higgs doublet
 with vacuum expectation value $v=174\GeV$:
\beq L =  \begin{pmatrix}\nu_\mu\cr \mu_L\end{pmatrix}, \qquad
E = \mu_R,\qquad
\qquad H =  \begin{pmatrix} 0\cr v+h/\sqrt{2}\end{pmatrix}.
\eeq 
In view of this connection,  we consider scenarios where both the operators
\beq {\cal O}_H = LEH^* + \hbox{h.c.}\quad\mathrm{and}\quad
{\cal O}_{HHH}=LEH^* (H^\dagger H)+\hbox{h.c.}\eeq contribute to the muon mass,
such that $m_\mu = m_\mu^H+m_\mu^{HHH}$: the muon mass can be due partially or totally to the new term.


Here we classify the  ``charged see-saw models'' where the new fermions 
mediate the operator ${\cal O}_{HHH}$ at tree level:
there are 6 different possibilities, illustrated in figure\fig{FeynTree}, and presented in the rest of this section.
Table~\ref{tab:listf} describes all new fermions present in such models.

\subsection{Charged see-saw with $L'$ and $E'$}
The first model (top-left Feynman diagram in figure\fig{FeynTree})
adds to the SM the following interactions between
left ($L$) and right-handed ($E$) muons with  
extra vector like lepton doublets $L'\oplus \bar L'$ 
\beq L' = \begin{pmatrix} L^{\prime 0} \cr L^{\prime -}\end{pmatrix} ,\qquad
\bar L' =  \begin{pmatrix} \bar L^{\prime 0}\cr \bar L^{\prime+}   \end{pmatrix},\eeq
and singlets $E'\oplus\bar E'$:
\beq -\Lag =  M_L \bar L' L' + M_E \bar E' E' +
 \lambda_L L'EH^* + \lambda_E LE'H^* + \bar\lambda_{LE} \bar L'\bar E'\ H +  \lambda_{LE} L'E' H^* +\hbox{h.c.}
\label{eq:La}
\eeq
Notice that this is the most general Lagrangian with the given field content,
that $L'$ and $E'$ 
have the same quantum numbers as $L$ and $E$,
and that $\bar L'$ and $\bar E'$ are independent fields whose quantum numbers are opposite to $L$ and $E$ respectively.

Integrating out at tree level the heavy  fermions gives rise to the following muon mass terms:
\beq \Lag_{\rm eff} = - \bigg[\lambda_\mu + \frac{\lambda_L \bar\lambda_{LE}\lambda_E }{M_L M_E} HH^\dagger\bigg] LEH^*+\hbox{h.c.}
=-(m_\mu^H +m_ \mu^{HHH}) \mu_L \mu_R+\hbox{h.c.}\label{eq:mmu}
\eeq
A mass mixing arises only in the charged lepton sector
\beq  \label{eq:ML'E'}\mathscr{M}_\pm=
\bordermatrix{ & \mu_R & \bar L^{\prime +} &E^{\prime +}\cr
\mu_L &  \lambda_\mu v & 0 & \lambda_Ev\cr
L^{\prime -} & \lambda_L v &  M_L &\lambda_{LE}v\cr
\bar E^{\prime -} & 0 & \bar\lambda_{LE} v & M_E} .
\eeq
Here and in the following we denote components of SU(2)$_L$ multiplets via their electric charge.

\subsection{Charged see-saw with $L_{3/2}$ and $E'$}
The middle-left model in figure\fig{FeynTree} employs $E'$ and $\bar E'$ (with mass term $M_E$)
and a doublet $L_{3/2}$ with hypercharge $Y =- 3/2$
together with the field $\bar L_{3/2}$ in the conjugated representation
and mass term $M_L$:
\beq \label{eq:L3/2}
L_{3/2} = \begin{pmatrix}L_{3/2}^- \cr L_{3/2}^{--}
\end{pmatrix} ,\qquad
\bar L_{3/2} =  \begin{pmatrix} \bar L_{3/2}^+ \cr \bar L_{3/2}^{++}\end{pmatrix}.\eeq
 The Yukawa Lagrangian is:
\beq-\Lag_{\rm Yuk} = 
 \lambda_L E (L_{3/2}\epsilon H) + \lambda_E LE'H^* + \bar\lambda_{LE}  \bar E'(\bar L_{3/2} \epsilon H^*) +  \lambda_{LE} E' (L_{3/2}\epsilon H) +\hbox{h.c.}
\label{eq:La:L32:E}
\eeq
where $\epsilon$ is the $2\times2$ asymmetric tensor.
The  mass mixing among leptons with $|Q|=1$ is like  in the previous model:
\beq  \label{eq:ML32E'}\mathscr{M}_\pm=
\bordermatrix{ & \mu_R & \bar L^{ +}_{3/2} &E^{\prime +}\cr
\mu_L &  \lambda_\mu v & 0 & \lambda_Ev\cr
L^{ -}_{3/2} & \lambda_L v &  M_L &\lambda_{LE}v\cr
\bar E^{\prime -} & 0 & \bar\lambda_{LE} v & M_E}, 
\eeq
consequently also the muon mass terms are  the same as in eq.\eq{mmu}.

\subsection{Charged see-saw with $L'$ and $E^a$}
The top-right model employs, in place of $E'$ and $\bar E'$, 
a weak triplet $E^a$ with hypercharge $Y=1$ plus the corresponding anti-triplet $\bar E^a$ with $Y=-1$:
\beq\label{eq:Ea}
E^a  \tau^a = 
\begin{pmatrix}
E^+ & \sqrt{2} E^{++} \\
\sqrt{2} E^0 & -E^+
\end{pmatrix}, 
\quad
\bar{E}^a  \tau^a = 
\begin{pmatrix}
\bar{E}^- & \sqrt{2} \bar{E}^{0} \\
\sqrt{2} \bar{E}^{--} & -\bar{E}^-
\end{pmatrix}.
\eeq
The Lagrangian has the same structure as in\eq{La}, with modified
gauge structure of the relevant interactions such as:
\beq  {\lambda_E}(H^*\tau^a L)E^a = \lambda_E \left(v+\frac{h}{\sqrt{2}} \right) (\sqrt{2} \nu_\mu E^0 - \mu_L E^+).
\eeq
Eq.\eq{mmu} remains valid, in view of 
$(H^* \tau^a L) (H^*\tau^a H) = (H^*L)(H^*H)$.
Non-trivial mass matrices arise now in both the charged and neutral sector:
\beq  \mathscr{M}_\pm = \bordermatrix{ & \mu_R & \bar L^{\prime +} & E^+\cr
\mu_L &  \lambda_\mu v & 0 &- \lambda_Ev\cr
L^{\prime -} & \lambda_L v &  M_L &-\lambda_{LE}v\cr
\bar E^- & 0 &- \bar\lambda_{LE} v & M_E}, \qquad
\mathscr{M}_0 = \bordermatrix{ &\bar\nu' & E^0 \cr
\nu_\mu &  0 & \sqrt{2}\lambda_Ev\cr
\nu' &   M_L & \sqrt{2}\lambda_{LE}v\cr
\bar E^0&   \sqrt{2}\bar\lambda_{LE} v & M_E} .
\eeq

\subsection{Charged see-saw with $L_{3/2}$ and $E^a$}
The middle-right model employs  the $L_{3/2}$ fields of
eq.\eq{L3/2}
 together with the $E^a$ and $\bar E^a$ fields of eq.\eq{Ea}.
The Lagrangian is again analogous to\eq{La}, with modified
gauge structure of the relevant interactions:
\begin{align}
\lambda_{LE}(L_{3/2}\epsilon\tau^a H)E^a &=\lambda_{LE}
\left( v+\frac{h}{\sqrt{2}} \right) (-\sqrt{2}  L_{3/2}^{--} E^{++} - L_{3/2}^- E^+),
\\
\bar\lambda_{LE}(H^*\tau^a \epsilon\bar L_{3/2})\bar E^a &= \bar\lambda_{LE}
\left( v+\frac{h}{\sqrt{2}} \right) (\sqrt{2} \bar L_{3/2}^{++} \bar E^{--} + \bar L_{3/2}^+ \bar E^-)
\end{align}
and lead to a mass mixing for charged as well as for doubly charged fields:
\beq  \mathscr{M}_\pm = \bordermatrix{ & \mu_R & \bar L_{3/2}^+ &  E^+ \cr
\mu_L &  \lambda_\mu v & 0 &- \lambda_Ev \cr
 L_{3/2}^- & \lambda_L v &  M_L &-\lambda_{LE}v \cr
\bar E^- & 0 & \bar\lambda_{LE} v & M_E}, \qquad
\qquad
\mathscr{M}_{\pm\pm} = \bordermatrix{ & \bar L_{3/2}^{++} &  E^{++} \cr
L_{3/2}^{--} & M_L & -\sqrt{2}\lambda_{LE}v   \cr
\bar E^{--}  & \sqrt{2}\bar\lambda_{LE}v & M_E   } \qquad
\eeq
and consequently again to eq.\eq{mmu} for $m_\mu^{HHH}$.

\subsection{Charged see-saw with $L'$ and $N'$}
The model in bottom-right position employs a $L'$ together with a right-handed neutrino $N'$ (like in type-I see-saw models \cite{Minkowski:1977sc,Yanagida:1979uq,Gell-Mann:1979kx,Glashow:1979nm,Mohapatra:1979ia}).
It has two problems:
\begin{itemize} \item[i)]  It generates the muon mass
operator $(LH)E(H^\dagger H^*)$ which vanishes due to the anti-symmetric SU(2) contraction in the second term.  No muon mass term is generated, but still the model can generate at one-loop level a muon magnetic moment.

\item[ii)] It generates neutrino masses.  This can be prevented by assuming two right-handed neutrinos $N'$ and $\bar N'$
with Dirac mass term $M_N \bar N' N'+\hbox{h.c}$. 
Unlike in the previous models this structure is not demanded by electroweak gauge invariance:
it can be obtained demanding conservation of lepton number, which is no longer an accidental symmetry.
\end{itemize}
Thus we consider the following lepton-number conserving
Lagrangian:
\beq -\Lag =  M_L \bar L' L' + M_N \bar N' N' +
 \lambda_L L'EH^* + {\lambda_N} L\epsilon H N' + {\bar\lambda_{LN}} \bar L' \epsilon H^*\bar N'
 +\lambda_{LN} L'\epsilon HN'
   +\hbox{h.c.}
\label{eq:LN}
\eeq
that gives rise to the following mass mixing only among neutral fermions
\beq 
\mathscr{M}_0= \bordermatrix{ &\bar L^{\prime 0}&  N^{\prime 0}\cr
\nu_\mu &  0 & \lambda_Nv\cr
L^{\prime 0} &   M_L &\lambda_{LN}v\cr
\bar N^{\prime 0} &  \bar\lambda_{LN} v & M_N} .
\eeq
and consequently does not yields any correction to the muon mass: $m_\mu^{HHH}=0$.

\subsection{Charged see-saw with $L'$ and $N^a$}
Finally, the bottom-right model in figure\fig{FeynTree}
employs $L'$ together with $N^a$, an SU(2) triplet with zero hypercharge (like in type-III see-saw models \cite{Foot:1988aq,Ma:1998dn,Ma:2002pf}).
As in the previous model, it is necessary to add $\bar N^a$ and conservation of lepton number
obtaining Dirac mass terms $M_L \bar L' L' + M_N \bar N^a N^a$ and Yukawa couplings:
\beq -\Lag_{\rm Yuk} =  
 \lambda_L L'EH^* + {\lambda_N} (L\epsilon\tau^a H) N^a + {\bar\lambda_{LN}}(\bar L'\tau^a\epsilon H^*)\bar N^a  + {\lambda_{LN}} (L'\epsilon\tau^a H) N^a + \hbox{h.c.}
\label{eq:LNa}
\eeq
Decomposing into components
\beq \tau^a N^a = \begin{pmatrix}
N^0 &\sqrt{2} N^+\cr \sqrt{2} N^-  & -N^0\end{pmatrix},\qquad
 \tau^a \bar N^a = \begin{pmatrix}
\bar N^0 &\sqrt{2} \bar N^+\cr \sqrt{2} \bar N^-  & -\bar N^0\end{pmatrix}
\eeq
gives the following charged and neutral mass matrices:
\beq  \label{eq:ML'Na}\mathscr{M}_\pm=
\bordermatrix{ & \mu_R & \bar L^{\prime +} &N^+ &\cr
\mu_L &  \lambda_\mu v & 0 & -\sqrt{2}\lambda_N v \cr
L^{\prime -} & \lambda_L v &  M_L &- \sqrt{2}\lambda_{LN}  v\cr
\bar  N^-& 0 & \sqrt{2}\bar\lambda_{LN} v & M_N},\qquad
\mathscr{M}_0 = \bordermatrix{ &\bar L^{\prime 0} & N^0 \cr
\nu_\mu &  0 & -\lambda_Nv\cr
L^{\prime 0} &   M_L &-\lambda_{LN}  v \cr
\bar N^0&  \bar\lambda_{LN} v & M_N}.
\eeq
The resulting see-saw contribution to the muon mass is $m_\mu^{HHH} =-2 v^3 \lambda_L \bar\lambda_{LN}\lambda_N/M_L M_N$.
The  states $\bar N^+$ and $N^-$ have been omitted from the mass matrix $\mathscr{M}_\pm$ because they
have opposite lepton number with respect to the other states
with the same charge, so they do not mix with the other states and will play no r\^{o}le in the following.

\section{Computing the muon anomalous magnetic moment}\label{g-2}

This section describes the technical details of the computation of the one-loop contribution to the anomalous magnetic moment,
fully determined in terms of the mass matrices for the charged states and 
the neutral states listed in the previous section, together with their gauge couplings.
The main points will be summarized in the next section, where we will present the results.

\subsection{Gauge interactions}
Here we list the explicit values of the gauge couplings of the various fields present in all 6 see-saw models needed for computing their
contribution to the muon $a_\mu$.


We start with the  interactions  with the $Z$ boson of fields with electric charge $|Q|=1$. 
We describe such fields, generically denoted as  $ \chi_i^-$ and $ \chi_i^+$, in terms of left-handed two-component Weyl spinors in the flavor eigenstate basis,
and write their couplings in the usual way:
\begin{equation}
\Lag \supset -\frac{g_2}{c_{\rm W}}\sum_i\left[
  g_{Zi}^L\,( \chi_i^-)^\dagger\bar\sigma^\mu \chi_i^-
-
  g_{Zi}^R\,( \chi_i^+)^\dagger\bar\sigma^\mu \chi_i^+
\right]Z_\mu
\,.
\end{equation}
where $c_{\rm W} = \cos\theta_{\rm W}$, $s_{\rm W} = \sin\theta_{\rm W}$, $\theta_{\rm W}$ is the weak angle, and
the explicit values for the couplings are
\beq \begin{array}{c|cccccccccc}
Q=-1& \mu_L & L^{\prime -} & \bar E^{\prime -} & E^-  & \bar N^-  & N^-  & L_{3/2}^- \cr
g_Z^L&  -{1\over2}+s_{\rm W}^2&-{1\over2}+s_{\rm W}^2 & s_{\rm W}^2 &s_{\rm W}^2 & -1+s_{\rm W}^2 & -1+s_{\rm W}^2 & {1\over2}+s_{\rm W}^2 \\[3mm] \hline \\[-3mm]
Q=+1& \mu_R & \bar L^{\prime +} &  E^{\prime +} & E^+  & \bar N^+  & N^+  & \bar L_{3/2}^+ \cr
g_Z ^R& s_{\rm W}^2 &  -{1\over2}+s_{\rm W}^2 & s_{\rm W}^2 &s_{\rm W}^2 & -1+s_{\rm W}^2 & -1+s_{\rm W}^2 & {1\over2}+s_{\rm W}^2 \\[4mm]
\end{array}
\eeq
Next, we need the $W$ couplings between neutral fields $N_i$
and charged fields $\chi^\pm_i$,
as well as between charged and doubly charged fields $\rho_i^{\pm\pm}$. 
We write them again in terms of left-handed Weyl spinors in the flavor basis $  N_i, \chi_i^\pm, \rho_i^{\pm\pm}$ as
\begin{align}
\Lag \supset
&-\frac{g_2}{\sqrt{2}} \sum_i\left[
  g_{W\!N\,i}^L\,(  N_i)^\dagger\bar\sigma^\mu \chi_i^-
-
  g_{W\!N\,i}^R\,( \chi_i^+)^\dagger\bar\sigma^\mu  N_i
\right]W_\mu^+ + \text{ h.c.}
\\
&-\frac{g_2}{\sqrt{2}} \sum_i\left[
  g_{W\!\rho\,i}^L\,( \rho_i^{--})^\dagger\bar\sigma^\mu \chi_i^-
-
  g_{W\!\rho\,i}^R\,( \chi_i^+)^\dagger\bar\sigma^\mu \rho_i^{++}
\right]W_\mu^- + \text{ h.c.}  \nonumber
\end{align}
Here we list  the non-zero couplings among the multiplets present in all the charged see-saw models:
\renewcommand{\arraystretch}{1.1}
\beq \begin{array}{c|cccccccccc}
Q=-2&  &   & \bar E^{--}  &  &  & L_{3/2}^{--}\cr
Q=-1& \mu_L & L^{\prime -}  & \bar E^-  & \bar N^-  & N^- & L_{3/2}^- \cr
Q=0& \nu_\mu & \nu'  & \bar E^0  & \bar N^0 & N^0 &  \cr \hline
g_{W\!N}^L& 1 & 1 & -\sqrt{2} & \sqrt{2} & \sqrt{2} \cr
g_{W\!\rho}^L&  & & \phantom{-}\sqrt{2} & & & 1\cr
\end{array}
\qquad~
\begin{array}{c|cccccccccc}
Q=+2&    & E^{++}  &  &  & \bar L_{3/2}^{++}\cr
Q=+1&   \bar L^{\prime +}  & E^+  & \bar N^+  & N^+ & \bar L_{3/2}^+ \cr
Q=0&   \bar\nu'  & E^0  & \bar N^0 & N^0 &  \cr \hline
g_{W\!N}^R& 1 & -\sqrt{2} & \sqrt{2} & \sqrt{2} \cr
g_{W\!\rho}^R& & \phantom{-}\sqrt{2}& & &1 \cr
\end{array}\eeq

\renewcommand{\arraystretch}{1.5}
\begin{table}[t]
\footnotesize
\begin{center}
\begin{tabular}{cccc}
Interaction  &Mass&  $\hat g^L_Z$ & $\hat g_Z^R$
\\\hline
$\mu Z \mu$ &$m_\mu$&   $\displaystyle-\frac{1}{2}+s_{\rm W}^2+v^2\frac{\lambda_E^2}{2M_E^2}$ & $\displaystyle s_{\rm W}^2-v^2\frac{\lambda_L^2}{2M_L^2}$ \\
$\mu Z \chi_1$ & $m_{\chi_1}\approx M_{L}$&   $\displaystyle v^2\frac{\lambda_E (M_E \lambda_{LE} + M_L \bar\lambda_{LE})}{2M_E(M_E^2-M_L^2)}$ & $ \displaystyle v\frac{\lambda_L}{2M_L}$ \\
$\mu Z \chi_2$ & $m_{\chi_2}\approx  M_{E}$&  $\displaystyle -v\frac{\lambda_E}{2M_E}$ & $\displaystyle v^2\frac{\lambda_L (M_L \lambda_{LE} + M_E \bar\lambda_{LE})}{2M_L(M_E^2-M_L^2)}$ \\[3mm]  \hline
Interaction  &Mass&$\hat g_{W\!N}^L$ & $\hat g_{W\!N}^R$ \\ \hline
$\mu W \nu_\mu$ &$m_{\nu_\mu}\approx 0$ & $ \displaystyle 1-v^2\frac{\lambda_{E}^2}{2M_E^2}$ & 0  \\
$\mu W N$ &$m_{N}\approx M_L$& $\displaystyle  v^2\frac{M_L\bar\lambda_{LE}\lambda_E-M_E\lambda_{\mu} \lambda_L}{M_EM_L^2}$ & $\displaystyle{-}v\frac{\lambda_L}{M_L}$  
\\[5mm] \hline
Interaction  &Mass&$\hat y_L^*$ & $\hat y_R$ \\ \hline
$\mu h \mu$ &$m_\mu$& $\lambda_\mu$ & $\lambda_\mu$  \\
$\mu h \chi_1$ & $m_{\chi_1}\approx M_{L}$& $v \left(\displaystyle
-\frac{\lambda_E \bar{\lambda}_{LE}}{M_E}
 - \lambda_E\frac{M_L \lambda_{LE}^* + M_E \bar{\lambda}_{LE}}{M_E^2-M_L^2} \right)$
& $\lambda_L$   \\
$\mu h \chi_2$ & $m_{\chi_2}\approx M_{E}$&  $\lambda_E$  &  
$\displaystyle v \left(   
-\frac{\lambda_L \bar{\lambda}_{LE}}{M_L}
+\lambda_L \frac{M_E \lambda_{LE} + M_L \bar{\lambda}_{LE}}{M_E^2-M_L^2}\right)$  
\\
\end{tabular}
\end{center}
\caption{ Couplings of muons to other fermions and to the $Z,W,h$ bosons in the $L'\oplus E'$ model. Without loss of generality, we have set the phases of the masses and couplings to zero, with the exception of $\lambda_{LE}$.
}
\label{tab:VZW}
\end{table} \renewcommand{\arraystretch}{1}

\subsection{Mass eigenstates}
Next, we diagonalize the mass matrices finding the neutral, charged and doubly-charged mass eigenstates, $\hat N_i$, $\hat \chi_i$ and $\hat \rho_i$ 
respectively in terms of the interaction eigenstates $N_i, \chi_i, \rho_i$:
\beq \begin{array}{l}
\mathscr{M}_0= L^\dagger_0 \cdot \diag (0, m_{N_1}, m_{N_2},\ldots ,)\cdot R_0,\\[2mm]
\mathscr{M}_{\pm} = L^\dagger_- \cdot \diag (m_\mu, m_{\chi_1},m_{\chi_2},\ldots )\cdot R_+,\\[2mm]
\mathscr{M}_{\pm\pm} = L^\dagger_{--} \cdot \diag ( m_{\rho_1},m_{\rho_2},\ldots )\cdot R_{++}\ .
\end{array}
\eeq
Switching to a four-component notation, grouping the fields into neutral, charged and doubly charged Dirac spinors, the gauge boson couplings in the mass eigenstate basis become
\begin{align}
\Lag \supset
&-\frac{g_2}{c_W}\sum_i
\hat{\bar\chi}_i\gamma^\mu \left(
\hat g_{Z\,ij}^L \, P_L + \hat g_{Z\,ij}^R \, P_R
\right)\hat\chi_j ~ Z_\mu
\\
&-\frac{g_2}{\sqrt{2}}\sum_i
\hat{\bar N}_i\gamma^\mu \left(
\hat g_{W\!N\,ij}^L \, P_L + \hat g_{W\!N\,ij}^R \, P_R
\right)\hat\chi_j ~ W_\mu^+ + \text{ h.c.}
\\
&-\frac{g_2}{\sqrt{2}}\sum_i
\hat{\bar \rho}_i\gamma^\mu \left(
\hat g_{W\!\rho\,ij}^L \, P_L + \hat g_{W\!\rho\,ij}^R \, P_R
\right)\hat\chi_j ~ W_\mu^- + \text{ h.c.}
\,,  \
\end{align}
where
\begin{align}
\hat g_{Z\,ij}^L &= (L_-)_{ik} \,   g_{Zk}^L \, (L_-)_{jk}^*
\,,&
\hat g_{Z\,ij}^R &= (R_+)_{ik} \,   g_{Zk}^{R*} \, (R_+)_{jk}^*
\,,\\
\hat g_{W\!N\,ij}^L &= \,(L_0)_{ik} \,   g_{Wk}^L \, (L_-)_{jk}^*
\,,&
\hat g_{W\!N\,ij}^R &= \,(R_0)_{ik} \,   g_{Wk}^{R*} \, (R_+)_{jk}^*
\,,\\
\hat g_{W\!\rho\,ij}^L &= \,(L_{--})_{ik} \,   g_{Wk}^L \, (L_-)_{jk}^*
\,,&
\hat g_{W\!\rho\,ij}^R &= \,(R_{++})_{ik} \,   g_{Wk}^{R*} \, (R_+)_{jk}^*
\,.
\end{align}
Although we will perform an exact numerical diagonalization we point out that 
analytical approximations can be obtained by performing a perturbative diagonalization of the mass matrices
in powers of their small off-diagonal entries: table~\ref{tab:VZW} summarizes the relevant resulting couplings
in the $L' \oplus E'$ model up to the order necessary to compute the leading $g-2$ effect.
Similar results apply to the other models, as all of them have similar mass matrices.

\begin{figure}[t]
\centering
\includegraphics[width=\textwidth]{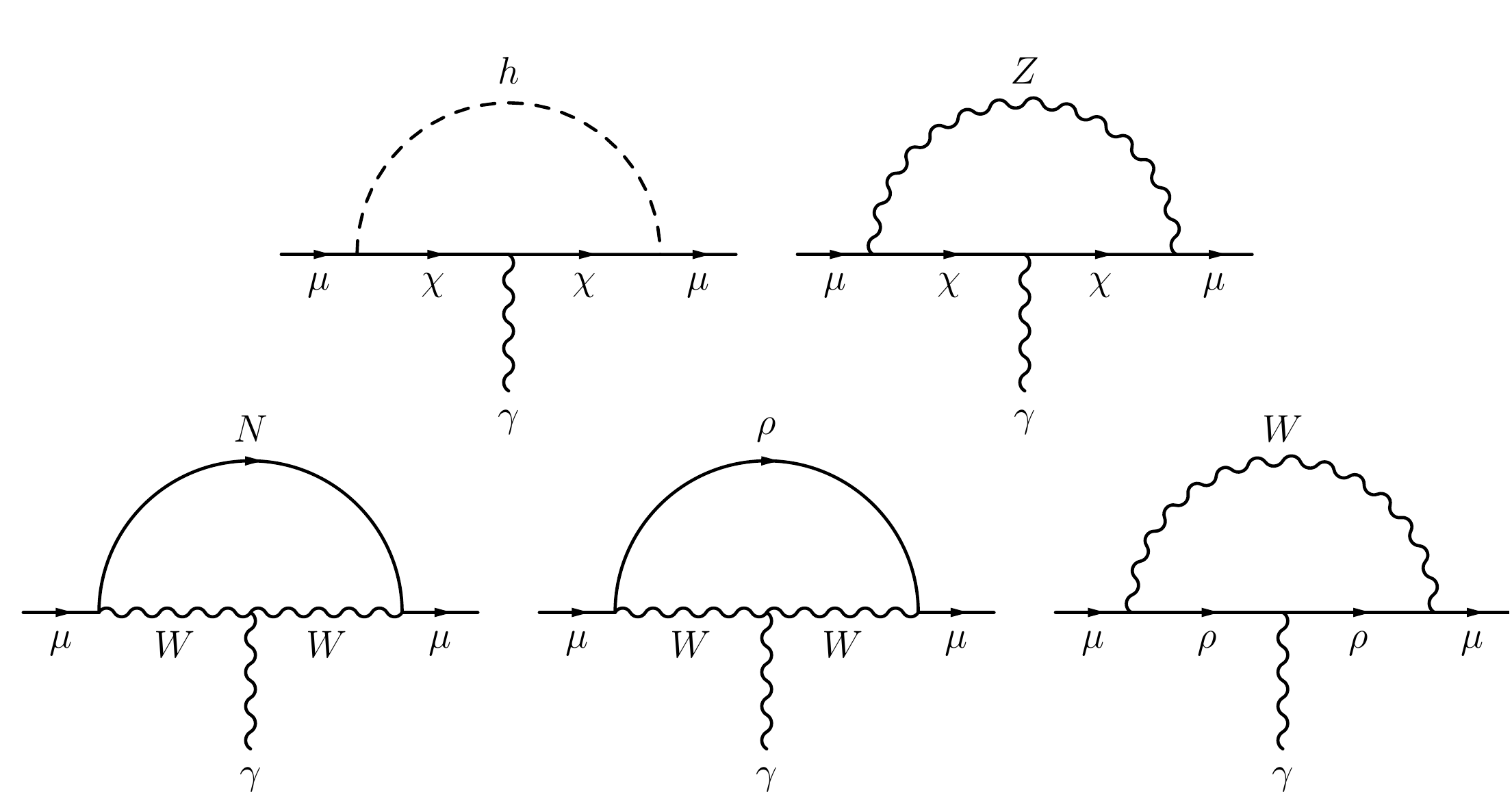}
\caption{ 1-loop Feynman diagrams with a new charged lepton $\chi$, a  neutral lepton $N$ and a doubly charged lepton $\rho$ that generate a muon magnetic moment.} 
\label{fig:1:loop:g-2}
\end{figure}

\subsection{The muon anomalous magnetic moment}
The one-loop electroweak contribution to the muon anomalous magnetic moment involving muons coupled to generic neutral fermions $N_j$, charged fermions $\chi_i$ and doubly charged fermions $\rho_k$ is given by the sum of the Higgs loop, $Z$ loop and $W$  loops, as illustrated in figure\fig{1:loop:g-2}:
\begin{equation}
\Delta a_\mu^\text{SM-EW}+\Delta a_\mu =
\sum_i \left[\delta a_\mu^h(\chi_i) +
\delta a_\mu^Z(\chi_i) \right]+
\sum_j \delta a_\mu^{W N}( N_j) +
\sum_k \delta a_\mu^{W\rho}(\rho_k).
\end{equation}
Notice that the SM electroweak contribution $\Delta a_\mu^\text{SM-EW}$ is included in the result
(because muons and their neutrinos with non-standard couplings to the Higgs are present among the charged and neutral fields) and that,
due to electromagnetic gauge invariance, at one loop the photon does not give any new physics contribution.
We neglect terms of order $m_\mu^2/M_{W,Z,h}^2$ and terms suppressed by $m_\mu/m_F$, where $m_F$ is a heavy fermion mass.
We now give explicit expressions for the various terms,  computed from the relevant diagrams in figure\fig{1:loop:g-2}.

\subsubsection[$Z$ contribution]{\boldmath $Z$ contribution}

The $Z$ loop contribution is
\begin{equation}\label{eq:aZ}
\delta a_\mu^Z = \frac{m_\mu^2}{(4\pi v)^2}
\left[
\left( |\hat g_Z^L|^2+|\hat g_Z^R|^2 \right) F_Z(x_Z)
+
\frac{m_\chi}{m_\mu}\text{Re}\left(\hat g_Z^L \hat g_Z^{R*}\right) G_Z(x_Z)
\right],
\end{equation}
where $x_Z=m_\chi^2/M_Z^2$ and
\begin{align}
F_Z(x) &= \frac{-5 x^4+14 x^3-39 x^2+18 x^2 \ln x+38 x-8}{3 (x-1)^4}
\,,\\
G_Z(x) &= \frac{2 \left(x^3+3 x-6 x \ln x-4\right)}{(x-1)^3}
\,.
\end{align}
\subsubsection[$W$ contribution with neutral fermion]{\boldmath $W$ contribution with neutral fermion}

The $W$ loop contribution is
\begin{equation}\label{eq:aW1}
\delta a_\mu^{W N} =\frac{m_\mu^2}{(4\pi v)^2}
\left[
\left( |\hat g_{W\!N}^L|^2+|\hat g_{W\!N}^R|^2 \right) F_{W  N}(x_{W  N})
+
\frac{m_{N}}{m_\mu}\text{Re}\left( \hat g_{W\!N}^L \hat g_{W\!N}^{R*} \right) G_{W  N}(x_{W  N})
\right], 
\end{equation}
where $x_{W  N}=m_{N}^2/m_W^2$ and\footnote{We thank Aditi Raval and Paride Paradisi for pointing out a mistake in $G_{W  N}(x)$ in the original version of this paper.}
\begin{align}
F_{W  N}(x) &= \frac{4 x^4-49 x^3+18 x^3 \ln (x)+78 x^2-43 x+10}{6 (x-1)^4}
\,,\\
G_{W  N}(x) &= \frac{4-15 x+12 x^2-x^3-6 x^2 \ln (x)}{(x-1)^3}
\,.
\end{align}

\subsubsection[$W$ contribution with doubly charged fermion]{\boldmath $W$ contribution with doubly charged fermion}

The $W$ loop contribution is
\begin{equation}\label{eq:aW2}
\delta a_\mu^{W\rho} =\frac{m_\mu^2}{(4\pi v)^2}
\left[
\left( |\hat g_{W\!\rho}^L|^2+|\hat g_{W\!\rho}^R|^2 \right) F_{W\rho}(x_{W\rho})
+
\frac{m_\rho}{m_\mu}\text{Re}\left( \hat g_{W\!\rho}^L \hat g_{W\!\rho}^{R*} \right) G_{W\rho}(x_{W\rho})
\right] 
\end{equation}
where $x_{W\rho}=m_\rho^2/m_W^2$ and
\begin{align}
F_{W\rho}(x) &= F_{W  N}(x) + F_Z(x)
\,,\\
G_{W\rho}(x) &= G_{W  N}(x) + G_Z(x)
\,.
\end{align}


\subsubsection{Higgs contribution}
We write the  Lagrangian coupling between the Higgs $h$, the muon $\mu$ and a charged mass eigenstate $\chi$ as
\begin{equation}
 -\frac{1}{\sqrt{2}}h [\bar{\hat \chi} (\hat y_L P_L + \hat y_R P_R) \mu] + \text{h.c.}
\end{equation}
The Higgs couplings $\hat y_L$ and $\hat y_R$ can be  derived making the substitution $v \to v + h/\sqrt{2}$ in the mass matrices
and rotating to the mass eigenstates;
their explicit values in the $L', E'$ model are summarized in table~\ref{tab:VZW}.
Then, the Higgs loop contribution to the muon $g-2$ is given by 
\begin{equation}\label{eq:ah}
\delta a^{h}_\mu = -\frac{m_\mu^2}{32\pi^2m_h^2}  \left[ (\abs{\hat y_{L}}^{2} + \abs{\hat y_{R}}^{2})  F_h(x_h) 
+ \frac{m_{\chi}}{m_{\mu}} \mathrm{Re}(\hat y_{L}^{*} \hat y_{R}) \, G_h(x_h)
 \right],
\end{equation}
where $x_h = m_{\chi}^{2}/m_{h}^{2}$ and
\begin{align}
F_h(x) &= -\frac{2 + 3 x - 6 x^2 + x^3 + 6 x \ln x}{6 (x-1)^4}, \\
G_h(x) &= -\frac{3 - 4 x + x^2 + 2 \ln x}{(x-1)^3}.
\end{align}
In the SM the Higgs loop contribution to the muon $g-2$ is suppressed by extra powers of $m_\mu/m_h$ with respect to the $W$ and $Z$ contributions, but in general the $h,Z$ and $W$ contributions are comparable.

\section{Results}\label{results}
Before discussing results for the muon magnetic moment, we consider the bounds from precision data,
showing that they allow for a new physics contribution to the muon mass, $m_\mu^{HHH}$, as large as the muon mass.

\begin{table*}[t]
$$\begin{array}{rlll}
\multicolumn{3}{c}{\hbox{\Blue Dimension 6 operators}}&
\qquad\hbox{Effects on precision observables}\Black \\  \hline
\Ord_{HL}' &=&i(H^\dagger D_\alpha \tau^a H)(\bar{L}_\mu\gamma_\alpha \tau^a L_\mu) +\hbox{h.c.}&
\delta g_{\mu_L}=-v^2 c'_{HL}\qquad \delta g_{\nu_\mu}=+v^2 c'_{HL}\qquad \delta G_{\rm F}=2v^2 c'_{HL}\\
\Ord_{HL} &=&i (H^\dagger D_\alpha H)(\bar{L}_\mu\gamma_\alpha L_\mu) +\hbox{h.c.}&
\delta g_{\mu_L}=-v^2 c_{HL}\qquad \delta g_{\nu_\mu}=-v^2 c_{HL}\\
\Ord_{HE} &=& i (H^\dagger D_\alpha H)(\bar{E}_\mu\gamma_\alpha E_\mu) +\hbox{h.c.} &\delta g_{\mu_R}=+v^2 c_{HE}\\
\end{array}$$
\caption{  
Dimensions 6 operators affecting the electroweak precision tests,
with their contributions to the $Z$ couplings $g = T_3 - Q s_{\rm W}^2$
and to the Fermi constant measured from muon decay, $G_{\rm F}$.\label{tab:O}}
\end{table*}

\subsection{Bounds from precision data}\label{EWPT}


The mixing of $\mu$ and $\nu_\mu$ with extra charged and neutral states
modifies the gauge couplings of the mass eigenstates
$\hat \mu$ and $\hat \nu_\mu$ to the $W$ and $Z$ vectors.
Various electroweak precision observables are affected: the $\mu$ lifetime, the forward-backward and left-right
asymmetries involving muons; the $Z$ width into $\mu^+\mu^-$ and $\nu_\mu\bar\nu_\mu$.
All such effects can be described  integrating out at tree level the heavy fields obtaining
the following effective Lagrangian \cite{Buchmuller:1985jz,Barbieri:1999tm,Cacciapaglia:2006pk}:
\beq \Lag = \Lag_{\rm SM}  + c_{HL} {\cal O}_{HL} + c'_{HL} {\cal O}'_{HL}+c_{HE}{\cal O}_{HE},\eeq
where the three relevant operators and their effects are listed in table~\ref{tab:O}
(in this paper we restrict the effect to muons).
By performing a global fit of precision data, including LEP2~\cite{Barbieri:1999tm,Cacciapaglia:2006pk},  we find the following best-fit values:
\beq\begin{array}{l}
c_{HE} = (-2.1\pm1.0)~10^{-3},\\
c_{HL} = (+2.1\pm0.7)~10^{-3},\\
c'_{HL} = (0.12\pm0.23)~10^{-3},
\end{array}\qquad
\rho = \begin{pmatrix}1&-0.82&-0.22\cr -0.82 & 1 & 0.12\cr -0.22 & 0.12&1\end{pmatrix},
\eeq
where $\rho$ is the correlation matrix.
The various particles and couplings we considered contribute as:
\beq\begin{array}{cccc}
&  c_{HE} & c_{HL} & c'_{HL}\\
L' & -\lambda_L^2/2M_L^2 & 0 &0\\
L_{3/2} &  +\lambda_L^2/2M_L^2 & 0 &0\\
E' & 0 & -\lambda_E^2/4M_E^2 & -\lambda_E^2/4M_E^2 \\
E^a & 0 & -3\lambda_E^2/4M_E^2 & +\lambda_E^2/4M_E^2 \\
N' & 0 &+\lambda_N^2/4M_N^2 & -\lambda_N^2/4M_N^2  \\
N^a & 0 & +3\lambda_N^2/4M_N^2 & +\lambda_N^2/4M_N^2 
\end{array}\eeq
In each model the total effect is then given by the sum of the contributions of the
particles it employs. For example,
the $L' \oplus E'$ charged see-saw predicts
\beq c_{HL} = c'_{HL}=-\frac{\lambda_E^2}{4M_E^2},
\qquad
c_{HE}= - \frac{\lambda_L^2}{2M_L^2}
\eeq
such that we get the following bounds at $95\%$ C.L. (1 dof):
\beq \frac{v\lambda_E}{M_E} < 0.03,\qquad \frac{v\lambda_L}{M_L} < 0.04.\eeq
Inserting such bounds in\eq{mmu} we see that
 the contribution to the muon mass generated by the higher dimensional operator can be
as large as the observed muon mass:
\beq |m_\mu^{HHH}/m_\mu| \circa{<}1.5\bar\lambda_{LE}.\eeq
Assuming that all the various couplings are perturbative (for definiteness smaller than unity)
the heavy leptons must  be lighter than a few TeV in order to give a $m_\mu^{HHH}$ comparable to $m_\mu$.
Similar results apply in the other models.

\begin{figure}[t]
\centering
\includegraphics[width=\textwidth]{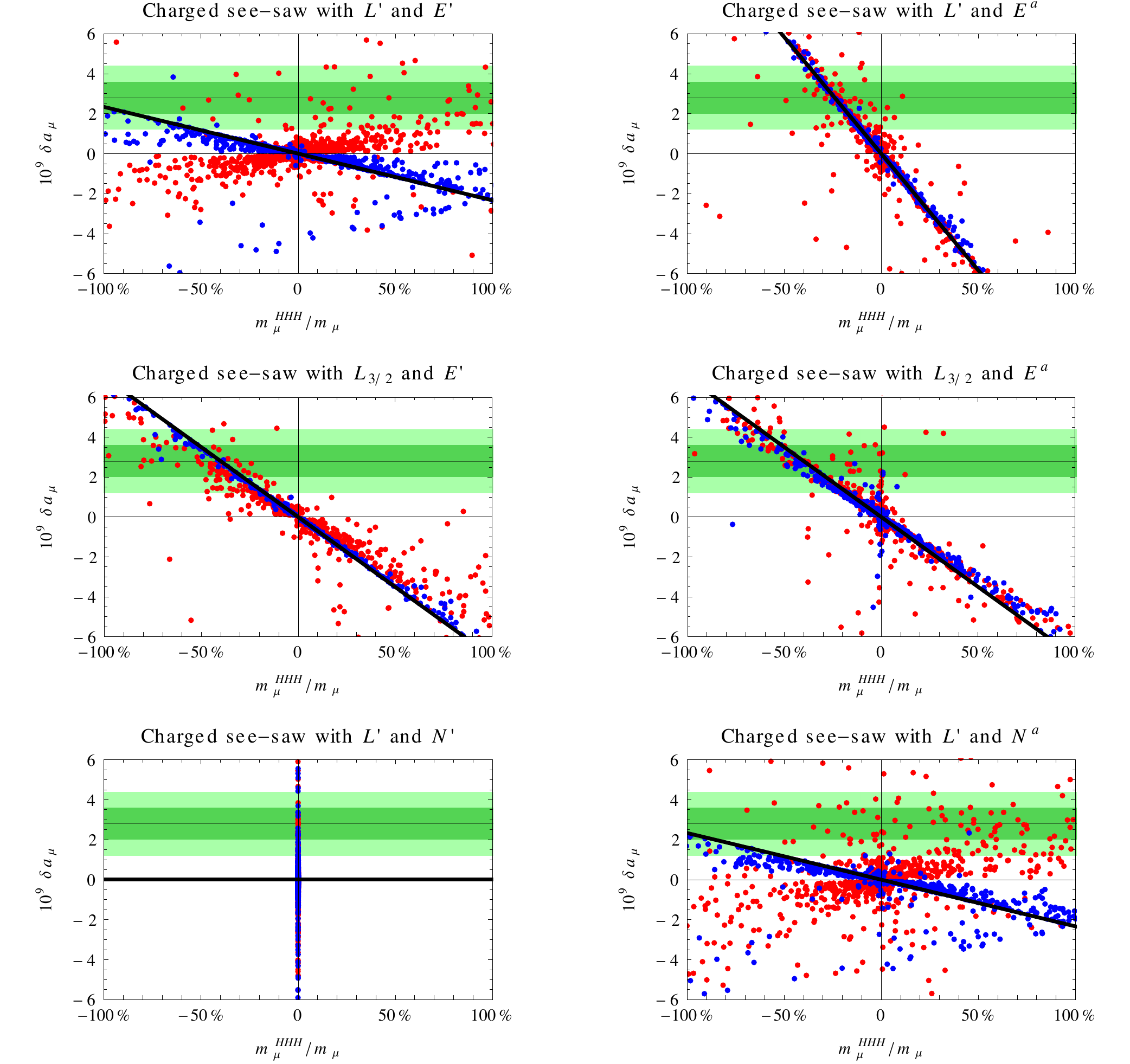}
\caption{ Random scan of the new-physics contribution to $a_\mu$ as function of the new-physics
percentage contribution to the muon mass, $m_\mu^{HHH}/m_\mu$;
red (blue) dots correspond to new leptons lighter (heavier) than $300 \GeV$.
We see that there is almost a one-to-one correspondence between $a_\mu$ and $m_\mu^{HHH}$,
in agreement with eq.\eq{amuapprox} (continuous black line).
The observed anomaly (horizontal green bands at $1\sigma$ and $2\sigma$)
is reproduced for $m_\mu^{HHH}/m_\mu\approx -1$ in the first $L'\oplus E'$ model,
and for similar values in all other models.
} 
\label{fig:scan}
\end{figure}

\subsection{The muon magnetic moment}
As the  RGE mixing between ${\cal O}_{HHH}$ and the muon magnetic operator happens to vanish,
one needs to consider the model-dependent one-loop corrections to the muon magnetic moment:
this technical computation was performed in section~\ref{g-2}.

The key difference with respect to similar computations in similar models with heavy leptons
which found negligible new-physics contributions to $a_\mu$~(see \cite{Biggio:2008in,Chua:2010me} and refs. therein) 
are the terms enhanced by the helicity flip on the heavy fermion mass, namely those terms multiplied by the functions $G$
in eqs.~\eq{aZ},\eq{aW1},\eq{aW2},\eq{ah}.
Such terms are present because we consider ``charged see-saw models'' where the heavy fermions give a new-physics contribution $m_\mu^{HHH}
\sim \lambda_L \bar\lambda_{LE}\lambda_E v^2/M_LM_E$
to the muon mass $m_\mu$,
which has the same chiral structure as the muon magnetic moment.
As a consequence these two new physics effects are expected to be strongly correlated:
\beq \Delta a_\mu \simeq c \frac{m_\mu m_\mu^{HHH}}{(4\pi v)^2} = 0.82 c  \frac{m_\mu^{HHH}}{m_\mu}\times \Delta a_\mu^{\rm exp},
\label{eq:amuapprox}
\eeq
where $c$ is an order one coefficient.
We can analytically compute both effects
in the limit in which the  new heavy leptons are degenerate and much heavier than the $W,Z$ bosons:
in the six charged see-saw models we find:

\beq \begin{array}{c|ccccccc}
c  &-1 &  -5 &  -3 &  -3 & \hbox{---} &-1\\[1mm]  \hline \\[-2ex]
\hbox{see-saw} & L'\oplus E' & L'\oplus E^a &L_{3/2}\oplus E' & L_{3/2}\oplus E^a & L'\oplus N' & L'\oplus N_a
\end{array}\eeq
This approximation does not apply to the $L'\oplus N'$ model, because in this case $m_\mu^{HHH}$
is accidentally zero, unlike $\Delta a_\mu$.


Figure\fig{scan} compares this approximation to generic scans in the six models of the new-physics contribution to the muon mass, $m_\mu^{HHH}$
versus the new-physics contribution to the muon anomalous magnetic moment, $\Delta a_\mu$.
The scan is restricted to values of parameters that satisfy the bounds from precision data discussed in section~\ref{EWPT},
and a full numeric computation of both $\Delta a_\mu$ and $m_\mu^{HHH}$ is performed.
We assumed a Higgs mass $m_h=120\GeV$.

\begin{itemize}
\item The blue dots, which correspond to new leptons heavier than $300\GeV$, lie along the lines corresponding to eq.\eq{amuapprox}.
In the first $L'\oplus E'$ model,  the new physics correction to $a_\mu$ equals to the observed anomaly for
$m_{\mu}^{HHH} \approx -m_\mu/3$.

\item
The red dots, which correspond to heavy leptons between $115\GeV$ and $300\GeV$, have a larger spread 
along the line corresponding to eq.\eq{amuapprox},
which comes from the dependence of the loop functions on the mass ratios.
\end{itemize}
The spread is so large that in most models the observed $a_\mu$ anomaly can be fitted without fine tunings
in the special case where the muon mass totally comes from new physics,
$m_\mu^{HHH} = m_\mu$, corresponding to the $+100\%$ point in figure\fig{scan}.

\begin{figure}[t]
\centering
\includegraphics[width=0.75\textwidth]{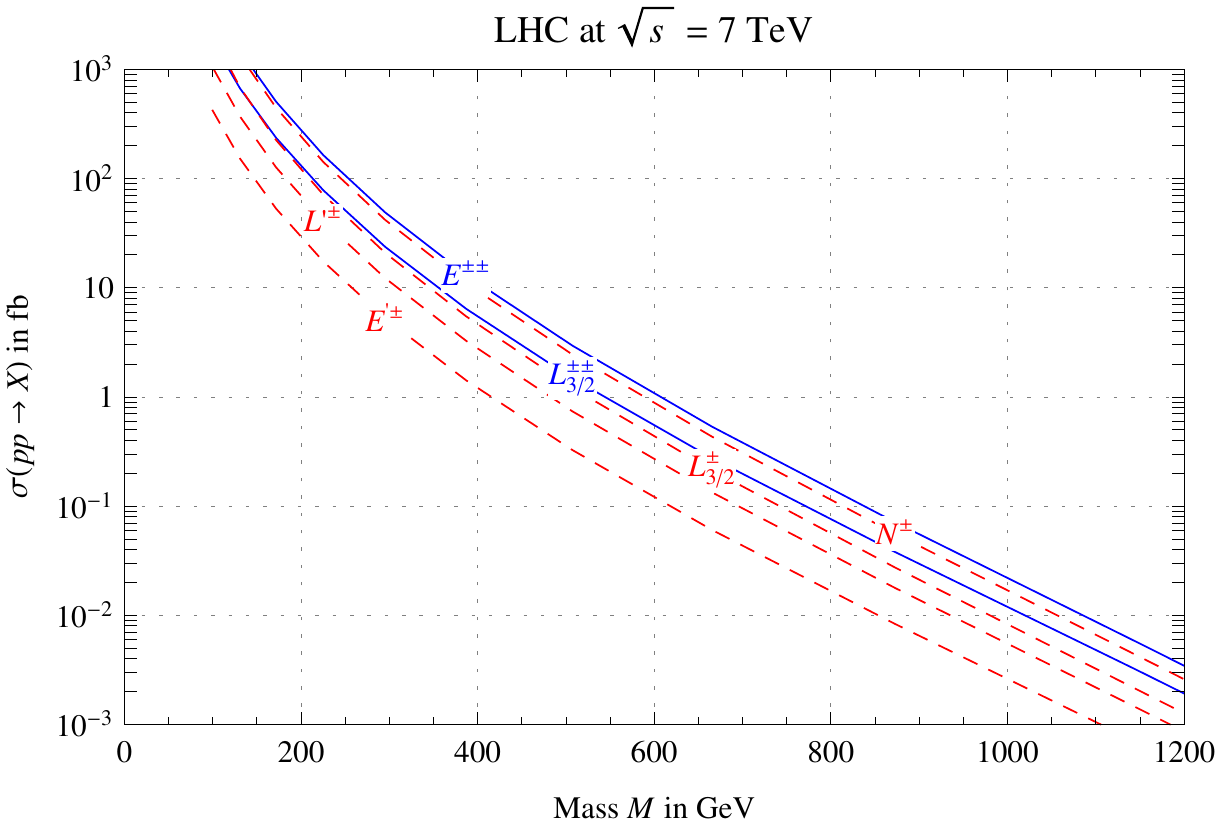}
\caption{ Cross sections for fermion-antifermion pair production at LHC.} 
\label{fig:LHC}
\end{figure}

\subsection{Signals at the LHC}

The negative searches at LEP imply that all the multiplets we consider must be heavier than 100 GeV,
with the only exception of $N'$ which has no gauge interactions.
The LHC can significantly extend the experimental sensitivity.

\subsubsection{Higgs decay rate into muons}
One observable model-independent consequence of the modified Higgs structure of the muon mass term
is the modified Higgs decay width into muons,
\beq\label{eq:h->mumu}
 \frac{\Gamma(h\to\mu^+\mu^-)}{\Gamma(h\to\mu^+\mu^-)_{\rm SM} } = \bigg| \frac{m_\mu^H +3m_\mu^{HHH}}{m_\mu}\bigg|^2\ ,
\eeq
which is enhanced by a factor of 9  if the muon mass is dominated by the new term, $|m_\mu^{HHH}|\gg |m_\mu^H|$.
Such a signal has been studied in~\cite{Giudice:2008uua}.

\subsubsection{Production rates of new heavy leptons}
Coming to the specific charged see-saw models, each one of them contains a particular set of heavy leptons with charges $|Q|=1$
accompanied by extra neutral or doubly-charged leptons.
The cross section for pair production is dominated by gauge interactions and fully predicted in terms of their quantum numbers.
Using the results from~\cite{DelNobile:2009st,Law:2011qe}, figure\fig{LHC} shows the predicted cross sections for pair production of the charged $\ell\bar\ell$ leptons present in the various see-saw models.
Production cross sections of particles with different charges $Q$ and $Q'$ are comparable.
We assumed the present LHC energy $\sqrt{s}=7\TeV$; for increased energies $s'$ the corresponding cross sections $\sigma_{s'}$ are obtained by rescaling the axes
as dictated by dimensional analysis:
\beq  \sigma_{s'}(m) = (s/s') \sigma_{s}(m\sqrt{s/s'}).\eeq
Single production is subdominant (unless the heavy leptons are very heavy) and will be neglected.

\subsubsection{Signatures of new heavy leptons}
One SU(2)$_L$ multiplet of heavy leptons corresponds to the scenario studied under the name of ``minimal matter''   in~\cite{DelNobile:2009st,Law:2011qe}:
the signals have the generic form
\beq pp \to V'V \ell \ell'\label{eq:MM},\eeq
where $\ell$ can be $\mu$ or $\nu_\mu$ and $V$ can be $W^\pm$ or $Z$ or $h$; each one of them has various decay modes.
So  the signal splits into many channels and the  event rate for each one of them is predicted in terms of the heavy lepton mass
(the predicted rates are typically comparable).
The situation is somehow analogous to the case of Higgs searches: combinations of various channels are needed to boost the sensitivity.
In particular the authors of~\cite{DelNobile:2009st,Law:2011qe} studied the main signals at LHC of the ``unusual''  $L_{3/2}$ and $E^a$ multiplets, together with the corresponding backgrounds:
$\ell^+\ell^+\ell^-\slashed{E}_T$ and $\ell^+\ell^+ \ell^-\slashed{E}_T jj$.

In the charged see-saw models one has at least two different new multiplets: the Yukawa interactions of ``minimal matter''
(heavy-lepton/lepton/Higgs) are supplemented by interactions among the new leptons and the Higgs in such a a way
that contributions to the muon mass and magnetic moment are generated.

So the signals typical of ``minimal matter'' in eq.\eq{MM} get supplemented by extra signals:
when the heavier multiplet is pair-produced, each one of them  now has two decay modes: 
\begin{itemize}
\item[a)] the one of ``minimal matter'' into $\ell V$;
\item[b)] into $L' V$ where $L'$ is the lighter heavy lepton, that subsequently decays into $\ell V$.
\end{itemize}
This means that the final states $\ell \ell' VV'$ of ``minimal matter''
can be supplemented by 0,1 or 2 extra $V$, obtaining the following final states:
\beq \ell\ell' VV,'\qquad \ell\ell' VV'V'',\qquad\ell\ell' VV'V''V'''  .\eeq
As far as we know no dedicated searches have been so far performed by the LHC collaborations, 
although some results about multi-leptons appeared in the context of other searches, see e.g. \cite{CMSPASEXO-11-045,CMS_PAS_SUS-11-013,cms3}.


\begin{figure}[t]
\centering
\includegraphics[width=0.5\textwidth]{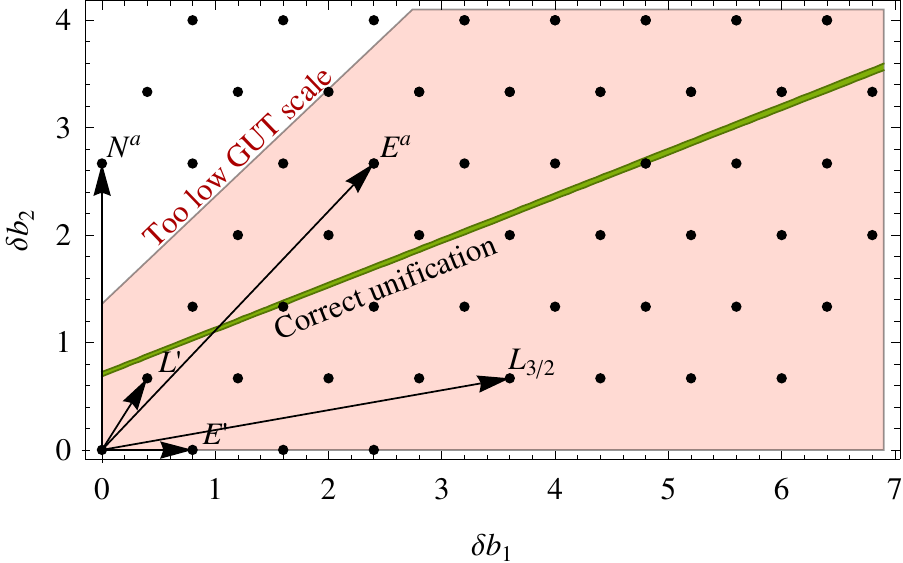}  
\caption{ Values of the new-physics contributions to the RGE coefficients
$b_1$ and $b_2$ for $\alpha_1$ and $\alpha_2$ produced by the heavy vector-like leptons present in the charged see-saw models
(arrows) and region favored by  gauge unification, assuming no new colored particles (within the green band the predicted $\alpha_3$
is within its $3\sigma$ uncertainty).
In the shaded region the unification scale is too low.
} 
\label{fig:plotGUT}
\end{figure}

\subsection{Renormalization up to higher energies}
Extrapolation up to higher energy of the  couplings in presence of the new heavy leptons at the weak scale
shows that a better unification of the gauge couplings can be achieved in some cases, while in others the unification is worse.
This is illustrated in figure\fig{plotGUT}, where, as function of generic new-physics contributions to the RGE coefficients
$b_1$ and $b_2$ for $\alpha_1$ and $\alpha_2$, we show the region compatible with the measured strong coupling constant~\cite{Bethke:2009jm}
\beq \alpha_3 (M_Z) = 0.1173\pm0.0007.\eeq
We ignored possible threshold corrections at the unification scale but considered 2-loop running effects.
The vectors in figure\fig{plotGUT} indicate the contributions to the RGE coefficients $(b_1,b_2$) produced by any single new vector-like lepton present in the ``charged see-saw models''.
For example in the $L'\oplus E'$  model one has to sum the vectors relative to $L'$ and $E'$.
The dots denote the sums obtained with $0,1,2,3$ copies of each heavy lepton.
Notice however that the unification scale, defined as the energy at which $\alpha_1 = \alpha_2$, is always below $10^{16}\GeV$,
and consequently the proton decay rate as predicted in SU(5) unification in 4 dimensions is too large
(unless its nucleon matrix element has been miscomputed~\cite{Martin:2011nd}).
As well known, new colored particles can avoid this problem but we will not explore this direction.

RGE equations for the various Yukawa couplings have the form:
\begin{eqnsystem}{sys:RGE}
(4\pi)^2 \frac{d\lambda_\mu}{d\ln \mu} &=& 
\lambda_\mu\bigg[X+\frac{5}{2}(\lambda_E^2+\lambda_L^2) + \lambda_{LE}^2+\bar\lambda_{LE}^2\bigg]+\frac{3}{2}\lambda_L \lambda_{LE}\lambda_E,\\
(4\pi)^2 \frac{d\lambda_{LE}}{d\ln \mu} &=&\lambda_{LE}
\bigg[X+\frac{5}{2}(\lambda_E^2+\lambda_L^2) + \lambda_{\mu}^2+\bar\lambda_{LE}^2\bigg]+ \frac{3}{2} \lambda_L \lambda_\mu \lambda_E,\\
(4\pi)^2 \frac{d\bar\lambda_{LE}}{d\ln \mu} &=& \bar\lambda_{LE}
\bigg[X+\frac{5}{2}\bar\lambda_{LE}^2+
\lambda_E^2+\lambda_L^2 + \lambda_{\mu}^2\bigg],\\
(4\pi)^2 \frac{d\lambda_{L}}{d\ln \mu} &=& \lambda_{L}
\bigg[X+\frac{5}{2}(\lambda_{L}^2+\lambda_{LE}^2+\lambda_\mu^2)+
\lambda_E^2+\bar\lambda_{LE}^2 \bigg]+\frac{3}{2}\lambda_E \lambda_{LE}\lambda_\mu, \\
(4\pi)^2 \frac{d\lambda_{E}}{d\ln \mu} &=& \lambda_{E}
\bigg[X+\frac{5}{2}(\lambda_{E}^2+\lambda_{LE}^2+\lambda_\mu^2)+
\lambda_L^2+\bar\lambda_{LE}^2 \bigg]+\frac{3}{2}\lambda_L \lambda_{LE}\lambda_\mu, 
\end{eqnsystem}
where $X = -\frac{9}{4} (g_1^2 +g_2^2) + 3 \lambda_{\rm top}^2$ and coefficients have been computed in the $L'\oplus E'$ model.
This means that the needed structure is stable under RGE.
The muon Yukawa coupling is generated at low energy even if it vanishes at the high-energy scale;
furthermore the new Yukawa terms in square brackets tend to reduce $\lambda_\mu$ at low energy.

The Yukawa couplings can now satisfy  the SU(5) relation $\lambda_\mu  = \lambda_s$ at the GUT scale, given the extra freedom
present in the model, both in the RGE and in the low-energy value of $\lambda_\mu$, no longer equal to $m_\mu/v$.

\section{Conclusions}\label{concl}
We classified the  6 ``charged see-saw models'' that can generate a contribution to the muon mass
and to the muon magnetic moment.
Such models only introduce new fermions around the weak scale: the motivation for
avoiding new scalars (and consequently new massive vectors) comes from the  interpretation 
of the smallness of the weak scale in terms of anthropic selection:   the Higgs is anthropically relevant and must be unnaturally light, 
while extra scalars would be anthropic irrelevant and should be naturally heavy.

All these ``charged see-saw models'' have the following structure: the SM fields $L$ (lepton doublet), $E$ (lepton singlet) and $H$ (Higgs doublet)
are supplemented by new heavy leptons $L'$ and $E'$ with Yukawa couplings
\beq L'EH^{*} + LE'H^{*} + L'E'H^{*},\eeq
such that $L$ and $E$ mix with $L'$ and $E'$ and the last Yukawa term becomes a contribution to the muon mass.
Indeed, integrating out the heavy states (see figure\fig{FeynTree}), one obtains the  dimension-6 effective operator $LEH |H|^2$,
which gives a contribution $m_\mu^{HHH}$ to the muon mass $m_\mu$.
Precision data allow for a muon mass totally produced by this new term.

As a consequence, the new physics contribution to the muon magnetic moment
\beq   \Delta a_\mu \sim \frac{m_\mu m_\mu^{HHH}}{(4\pi v)^2},\eeq
is  of the same order as the SM electroweak contribution
\beq  \Delta a_\mu^\text{SM-EW} \sim \frac{m_\mu^2}{(4\pi v)^2},
\eeq
and thus comparable to the discrepancy suggested by experiments,
see figure\fig{scan}.


We discussed  other signatures of this model: a modified Higgs/muon coupling, and the presence of
heavy leptons below a few TeV; possibly just above 100 GeV.

\acknowledgments 

This work was supported by the ESF grants 8090, 8499, 8943, MJD140, MTT8 and by SF0690030s09 project and by the European 
Union Programme ``Unification in the LHC Era",  contract PITN-GA-2009-237920 (UNILHC) and the European Regional Development Fund.


\bibliographystyle{JHEP} 
\bibliography{artv3}

\end{document}